\documentclass[10pt,a4paper]{article}
\usepackage{bm, amssymb,pifont,cancel, amsmath,comment,color,authblk}
\usepackage[dvips]{graphicx}
\usepackage{multicol}
\newcommand{\cW}{\mathcal{W}}
\newcommand{\cF}{\mathcal{F}}
\makeatletter

\newenvironment{figurehere}
  {\def\@captype{figure}}
  {}
\makeatother

\begin{document}
\begin{flushright}
WU-HEP-15-9\\
KEK-TH-1819
\end{flushright}
\begin{center}
{\Large{\bf{Massive vector multiplet inflation with Dirac-Born-Infeld type action}}}\\ 
\vskip 6pt
\large{Hiroyuki Abe}${}^1${\renewcommand{\thefootnote}{\fnsymbol{footnote}}\footnote[1]{E-mail address: abe@waseda.jp}}, \large{Yutaka Sakamura}${}^{2,3}${\renewcommand{\thefootnote}{\fnsymbol{footnote}}\footnote[2]{E-mail address: sakamura@post.kek.jp}} and \large{Yusuke Yamada}${}^1${\renewcommand{\thefootnote}{\fnsymbol{footnote}}\footnote[3]{E-mail address: yuusuke-yamada@asagi.waseda.jp}}\\
\vskip 4pt
${}^1${\small{\it Department of Physics, Waseda University,}}\\
{\small{\it Tokyo 169-8555, Japan}}\\
\vskip 1.0em

${}^2${\small\it KEK Theory Center, Institute of Particle and Nuclear Studies, 
KEK, \\ Tsukuba, Ibaraki 305-0801, Japan} \\ \vspace{1mm}
${}^3${\small\it Department of Particles and Nuclear Physics, \\
SOKENDAI (The Graduate University for Advanced Studies), \\
Tsukuba, Ibaraki 305-0801, Japan} 
\end{center}
\begin{abstract}
We investigate the inflation model with a massive vector multiplet in a case that the action of the vector multiplet is extended to the Dirac-Born-Infeld (DBI) type one. We show the massive DBI action in 4 dimensional ${\cal N}=1$ supergravity, and find that the higher order corrections associated with the DBI-extension make the scalar potential flat with a simple choice of the matter couplings. We also discuss the DBI-extension of the  new minimal Starobinsky model, and find that it is dual to a special class of the massive DBI action. 
\end{abstract}
\begin{multicols}{2}
\section{Introduction}\label{intro}
Cosmic inflation~\cite{Guth:1980zm,Starobinsky:1980te,Sato:1980yn} is a plausible solution for the problems of the standard Big Bang scenario, such as the flatness and the horizon problems. Especially, the slow-roll inflation models~\cite{Linde:1983gd} naturally provide the primordial density fluctuation, which is a source of the large scale structure of the universe. The primordial curvature perturbation is observed by the cosmic microwave background experiments, and their results support the slow-roll inflation models.

It is important to embed such inflation models to the UV complete theory, such as superstring theory or its effective theory, i.e., the supergravity (SUGRA). So far, many inflation models have been constructed in 4 dimensional (4D) ${\cal N}=1$ SUGRA. In many literatures, it is assumed that the inflationary potential is provided by the so-called F-term potential. However, there are some problems in such models due to the structure of SUGRA, such as the $\eta$ problem in SUGRA, which is caused by the factor $e^K$ in the F-term potential where $K$ denotes the K\"ahler potential. One of the possible solution for the problems was suggested in Ref.~\cite{Kawasaki:2000yn} and developed in Ref.~\cite{Kallosh:2010xz}. In those models, the existence of at least two chiral multiplets are required to realize inflation, and the multiplet other than the inflaton multiplet should have a sufficiently large quartic coupling in the K\"ahler potential, otherwise it becomes tachyonic 
or light, which may lead to an instability of the inflationary trajectory or an unacceptably large non-Gaussianity~\cite{Demozzi:2010aj}. The solution for such a problem is the realization of inflation without the additional chiral multiplet, that is, the inflation with a single multiplet. Such a model was first proposed in Ref.~\cite{Goncharov:1983mw}, and recently developed in Ref.~\cite{Ketov:2014qha}. 

The other interesting solution was suggested in Refs.~\cite{Kawano:2007gg,Ferrara:2013rsa,Farakos:2014gba}. In these models, the inflaton multiplet is a massive U(1) vector multiplet~\cite{VanProeyen:1979ks,Mukhi:1979wc}, which is equivalent to a combination of Stueckelberg chiral and anti-chiral multiplets and a massless vector multiplet. The inflation is driven by the scalar component of the massive vector multiplet. One of the advantages of such a model is that only the single massive vector multiplet is the sufficient source for inflation, and the $\eta$ problem is absent because the inflaton potential is provided by the so-called D-term potential. Therefore, the inflation is realized in a relatively simple way.

From the theoretical viewpoint, it is important to consider the UV completion of such SUGRA inflation models. Especially, in superstring theory, it is known that the action of vector fields are described by the Dirac-Born-Infeld (DBI) type action~\cite{Born:1934gh,Dirac:1962iy} if the vector fields are zero modes of open strings attached to the D-branes (see Ref.~\cite{Tseytlin:1999dj} for a review). Then, in such a case, the massive vector multiplet inflation may be realized with the DBI action. 

The supersymmetric (SUSY) realization of the DBI action has been intently studied e.g. in Refs.~\cite{Cecotti:1986gb,Bagger:1996wp,Rocek:1997hi,Ketov:1998ku}.\footnote{See also Refs.\cite{Tseytlin:1999dj,Ketov:2001dq} for reviews.} Especially, it was found that the DBI action in ${\cal N}=1$ 4D global SUSY naturally appears from the partial ${\cal N}=2$ SUSY breaking~\cite{Bagger:1996wp,Rocek:1997hi}, which is realized by imposing a condition between two ${\cal N}=1$ multiplets included in an ${\cal N}=2$ vector multiplet. The condition can also be embedded into SUGRA~\cite{Kuzenko:2002vk,Kuzenko:2005wh}, which realizes the DBI action in 4D ${\cal N}=1$ SUGRA. The matter coupled DBI action was also discussed in 4D ${\cal N}=1$ global SUSY~\cite{Ketov:2003gr} and SUGRA~\cite{Kuzenko:2002vk,Kuzenko:2005wh}. Recently, in SUGRA, we developed the matter coupled DBI type action including matters charged under the corresponding U(1) symmetry in Ref.~\cite{Abe:2015nxa}.

To explore the UV completions of the massive vector multiplet inflation models, we will extend the action of the massive vector multiplet to the DBI type action. Such an action can be constructed by the DBI action of a massless vector multiplet coupled to the Stueckelberg multiplet, which non-linearly transforms under the corresponding U(1) symmetry. The DBI action with charged matter multiplets was studied in our previous work~\cite{Abe:2015nxa} as mentioned above. Such a DBI-extended action contains higher order terms of the matter fields, associated with supersymmetric higher derivative couplings. We call them the DBI corrections throughout this paper. We found that the DBI corrections significantly modify  
the D-term potential. As we will show, the DBI corrections become important if the cut-off scale associated with the DBI action is smaller than the Planck scale. Contrary to a naive expectation, the corrections make the scalar potential flatter. This is an interesting feature of the DBI-extension. We demonstrate such a feature for two concrete models, which are the chaotic~\cite{Linde:1983gd} and the Starobinsky model~\cite{Starobinsky:1980te}. 

In Ref.~\cite{Cecotti:1987qe}, it is shown that the Starobinsky model
in the new minimal SUGRA is dual to  the massive vector multiplet action.
We will also investigate the DBI-extension of the former, and find that
it is dual to a special class of the massive DBI action,
which has a restricted form of the matter couplings.
Because of the restricted form, the flatness of the inflaton potential is
no longer protected under the DBI corrections.

The remaining parts of this paper are organized as follows. In Sec.~\ref{review}, we briefly review the massive vector action in SUGRA and its application to the inflation model based on Ref.~\cite{Ferrara:2013rsa}. Sec.~\ref{massiveDBI} is devoted to the construction of the DBI action of a massive vector multiplet. Then we obtain the modified D-term potential. In Sec.~\ref{inflation}, we discuss how the DBI corrections affect the inflaton potential, and find that the potential becomes flatter when the cut-off scale of the DBI sector is much smaller than the Planck scale. We also discuss the DBI-extension of the Starobinsky model in Sec.~\ref{DBIR^2}, and show the dual action of it. Finally, we conclude in Sec.~\ref{conclusion}. In Appendix~\ref{Starobinsky model}, we give a brief review of the Starobinsky model in the new minimal SUGRA.
\section{Review of the massive vector inflation}\label{review}
In this section, we briefly review the massive vector action in supergravity and the inflation models with it, based on Ref.~\cite{Ferrara:2013rsa}. The action in conformal SUGRA~\cite{Kaku:1978nz,Kugo:1982cu} is given by
\begin{align}
S=\left[\frac{1}{2}S_0\bar{S}_0\Phi(\Lambda+\bar{\Lambda}+gV)\right]_D-\frac{1}{4}[\cW^\alpha(V)\cW_\alpha(V)]_F\label{S1}
\end{align}
where $S_0$ is the chiral compensator, $\Lambda$ is a Stueckelberg chiral multiplet, $V$ is a vector multiplet, $g$ is the gauge coupling, $\Phi(x)$ is an arbitrary real function of $x$, $\cW^\alpha$ is the field strength supermultiplet, and $[\cdots]_{D,F}$ are the superconformal D- and F-term density formulae respectively~\cite{Kugo:1982cu}. The supergauge transformations of $\Lambda$ and $V$ are 
\begin{align}
\Lambda\rightarrow \Lambda+g\Sigma, \quad V\rightarrow V-\Sigma-\bar{\Sigma},\label{trans}
\end{align}
where $\Sigma$ denotes the gauge parameter chiral multiplet. By choosing $\Sigma=-\Lambda/g$, the vector multiplet $V$ becomes a massive vector multiplet\footnote{The action~(\ref{S1}) is also dual to the massive tensor action as shown in Refs.~\cite{Ferrara:2013rsa,Cecotti:1987qr}, however we do not discuss it further.}, and the action~(\ref{S1}) becomes 
\begin{align}
S=\left[\frac{1}{2}S_0\bar{S}_0\Phi(gV)\right]_D-\frac{1}{4}[\cW^\alpha(V)\cW_\alpha(V)]_F\label{S2}.
\end{align}

Let us focus on the bosonic part of the action~(\ref{S1}).\footnote{We can also regard the action~(\ref{S1}) as the ordinary supergravity action with the K\"ahler potential $K=-3\log (-\Phi(\Lambda+\bar{\Lambda}+gV)/3)$. }
For notational simplicity, we choose $\Phi(\Lambda+\bar{\Lambda}+gV)=-3e^{-2J/3}$, where $J=J(\frac{1}{2}(\Lambda+\bar{\Lambda}+gV))$. After imposing the conventional superconformal gauge conditions~\cite{Kugo:1982mr} and integrating out auxiliary fields, we obtain the following bosonic Lagrangian $\mathcal{L}_B$ in the Einstein frame,
\begin{align}
\mathcal{L}_B=&-\frac{1}{2}J''(C)\partial_\mu C\partial^\mu C-\frac{g^2}{2}J''\left(A_\mu-\frac{1}{g}\partial_\mu \theta\right)^2\nonumber\\
&-\frac{g^2}{2}(J'(C))^2-\frac{1}{4}\cF_{\mu\nu}\cF^{\mu\nu}+\frac{1}{2}R,\label{S3}
\end{align}
where $C={\rm Re} \Lambda$, $\theta={\rm Im}\Lambda$, the prime on $J$ denotes the derivatives with respect to $C$, $A_\mu$ is the vector component of $V$, $\cF_{\mu\nu}$ is the field strength of $A_\mu$, and $R$ denotes the Ricci scalar.\footnote{We use the Planck unit convention in which $M_{pl}\sim2.4\times10^{18}{\rm GeV}=1$.} By choosing the U(1) gauge as $\theta=0$, the second term in Eq.~(\ref{S3}) becomes the mass term of $A_\mu$ with $m_{A}^2=g^2J''(C)$.

In the following, we identify $C$ as the inflaton with its scalar potential given by
\begin{align}
V=\frac{g^2}{2}(J'(C))^2.\label{V}
\end{align}
We can reconstruct various types of the inflaton potential by choosing the function $J(C)$. For example, the choice $J=C^2/2$ gives the potential $V=g^2C^2/2$, which corresponds to the simplest version of the chaotic inflation~\cite{Linde:1983gd}.

Another interesting case is the Starobinsky model~\cite{Starobinsky:1980te} , which was first discussed in Ref.~\cite{Cecotti:1987qe} and applied to inflation in Ref.~\cite{Farakos:2013cqa}. In that case,  $J(C)$ takes a special form given by $J=-\frac{3}{2}\log( -Ce^C/3)$. (See Eq.~(42) in Appendix A.) The corresponding Lagrangian of $C$ is
\begin{align}
\mathcal{L}=-\frac{3}{4C^2}\partial_\mu C\partial^\mu C-\frac{9g^2}{8}(1+\frac{1}{C})^2.\label{R^21}
\end{align}
In terms of the canonical scalar $\phi=\sqrt{3/2}\log (-C)$, the Lagrangian~(\ref{R^21}) can be rewritten as
\begin{align}
\mathcal{L}=-\frac{1}{2}\partial_\mu \phi\partial^\mu \phi-\frac{9g^2}{8}(1-e^{-\sqrt{2/3}\phi})^2.\label{R^22}
\end{align}
\section{DBI action for a massive vector multiplet}\label{massiveDBI}
We extend the massive vector action reviewed in Sec.~\ref{review} to the DBI type one developed in our previous work~\cite{Abe:2015nxa}. Let us consider the following action,
\begin{align}
S=\left[\frac{1}{2}S_0\bar{S}_0\Phi(\Lambda+\bar{\Lambda}+gV)\right]_D-[hS_0^3X(\cW,\bar{\cW})]_F.\label{DBI1}
\end{align}
In the second term, $X(\cW,\bar{\cW})$ is a solution of the following equation,
\begin{align}
S_0^3X=\cW^\alpha\cW_\alpha-X\Sigma_c(\omega(\Lambda,\bar{\Lambda})\bar{S}_0\bar{X}),\label{cond}
\end{align}
where $X$ is a chiral multiplet, $\Sigma_c(\cdot)$ denotes the chiral projection operator in conformal SUGRA acting on the multiplet in its argument, $h$ is a real constant parameter, and $\omega(\Lambda,\bar{\Lambda})$ is a real function of $\Lambda$ and $\bar{\Lambda}$.

After the superconformal gauge fixings and integrating out the auxiliary fields, we obtain the following bosonic action,\footnote{For a detailed derivation of the action, see Ref.~\cite{Abe:2015nxa}.}
\begin{align}
S=&\int d^4x\sqrt{-g}\Biggl[\frac{1}{2}R-\frac{1}{2}J''(C)\partial_\mu C\partial^\mu C-\frac{g^2}{2}J''A_\mu A^\mu\nonumber\\
+\int& d^4x \frac{e^{\frac{4J}{3}}h}{\omega}\left(1-\sqrt{\tilde{P}}\sqrt{-{\rm det}\left(g_{\mu\nu}+2e^{-\frac{2J}{3}}\sqrt{\omega}\cF_{\mu\nu}\right)}\right),\label{DBI3}
\end{align}
where we have chosen the U(1) gauge condition as $\theta=0$, and 
\begin{align}
\tilde{P}\equiv1+\frac{\omega e^{-4J/3} g^2(J'(C))^2}{4h^2}.
\end{align}
In the following discussion, let us consider a simple case with $h=1/4$ and $\omega=\exp (4J/3)/(4M^4)$, where $M$ is a positive constant. Then the action~(\ref{DBI3}) becomes
\begin{align}
S=&\int d^4x\sqrt{-g}\Biggl[\frac{1}{2}R-\frac{1}{2}J''(C)\partial_\mu C\partial^\mu C-\frac{g^2}{2}J''A_\mu A^\mu\nonumber\\
+&\int d^4x M^4\left(1-\sqrt{P}\sqrt{-{\rm det}\left(g_{\mu\nu}+\frac{1}{M^2}\cF_{\mu\nu}\right)}\right),\label{DBI2}
\end{align}
where 
\begin{align}
P\equiv1+\frac{g^2(J'(C))^2}{M^4}.
\end{align}
In the limit of $M\rightarrow \infty$, the action~(\ref{DBI2}) reduces to the action~(\ref{S3}). Indeed, in this limit, we easily find
\begin{align}
S=&\int d^4x\left[-\frac{g^2(J'(C))^2}{2}-\frac{1}{4}\cF_{\mu\nu}\cF^{\mu\nu}+\mathcal{O}(M^{-4})+\cdots\right]
\end{align}
where the ellipsis denotes terms independent of $M$. We can also see this by noticing that, in such a limit, Eq.~(\ref{cond}) becomes $S_0^3X=\cW^\alpha\cW_\alpha$. Then, the action~(\ref{DBI1}) becomes exactly the same as one in Eq.~(\ref{S1}).

Therefore, we can regard the action~(\ref{DBI2}) as the DBI type extension of a massive vector multiplet, and $M$ as a cut-off scale of the DBI sector. Note that the massive DBI action loses the gauge invariance which the original one~\cite{Born:1934gh} has, due to the spontaneous breaking of the U(1) symmetry. Such a spontaneous breaking occurs, e.g. in superstring models when the Green-Schwartz anomaly cancellation~\cite{Green:1984sg} is realized, which can be described by the Stueckelberg multiplet as in Ref.~\cite{Lopes Cardoso:1991zt}.
\section{Inflaton potential}\label{inflation}
In the previous section, we derived the DBI action of a massive vector multiplet. In this section, let us discuss the effects of the DBI corrections on the scalar potential, and its implication for inflation models. 

From Eq.~(\ref{DBI2}), we obtain the DBI-extended scalar potential of $C$,
\begin{align}
V_{\rm mod}=&M^4(\sqrt{P}-1)\nonumber\\
=&M^4\left(\sqrt{1+\frac{g^2(J'(C))^2}{M^4}}-1\right).\label{modV}
\end{align}  
In Ref.~\cite{Ferrara:2013kca}, it is concluded that in inflation models with a massive vector multiplet, the higher order corrections associated with SUSY higher derivative terms are negligibly small in general. Indeed, all the DBI corrections in Eq.~(\ref{modV}) are proportional to the coupling constant $g^{2n}/M^{4(n-1)}$ ($n\geq 2$). Then, they are negligible compared to the leading term $g^2(J'(C))^2/2$ because $g$ should be much smaller than 1 to explain the amplitude of the primordial density perturbation.

However, we should notice that, even in the case of $g\ll1$, the higher order corrections become important if $M$ is also much smaller than 1. In fact, a parameter that controls the DBI corrections is $\beta\equiv g^2/M^4$, rather than $g^2$. In terms of $\beta$, we rewrite the scalar potential~(\ref{modV}) as
\begin{align}
V_{\rm mod}=\frac{g^2}{\beta}(\sqrt{1+\beta (J'(C))^2}-1).\label{modV2}
\end{align}
We find that for sufficiently large $\beta$ the scalar potential is approximately given by $V\sim (g^2/\sqrt{\beta})|J'(C)|$. 

In general, when the cut-off scale $M$ is low, the higher order corrections ruin the flat profile of the scalar potential. Contrary to such a naive expectation, we notice that in the regime of $\beta\gg1$, the DBI corrections make the shape of the scalar potential flatter than that of the ordinary D-term potential~(\ref{V}), given by $V\sim (J'(C))^2$. 

As a concrete example,  let us consider the case with $J=C^2/2$, which corresponds to the simplest chaotic inflation model mentioned in Sec.~\ref{review}. Then the scalar potential is given by
\begin{align}
V=\frac{g^2}{\beta}(\sqrt{1+\beta C^2}-1).\label{modV3}
\end{align}
We show the form of the scalar potential with various values of $\beta$ in Fig.~\ref{chaotic.fig}. 
\begin{figurehere}
\begin{center}
\includegraphics[width=85mm]{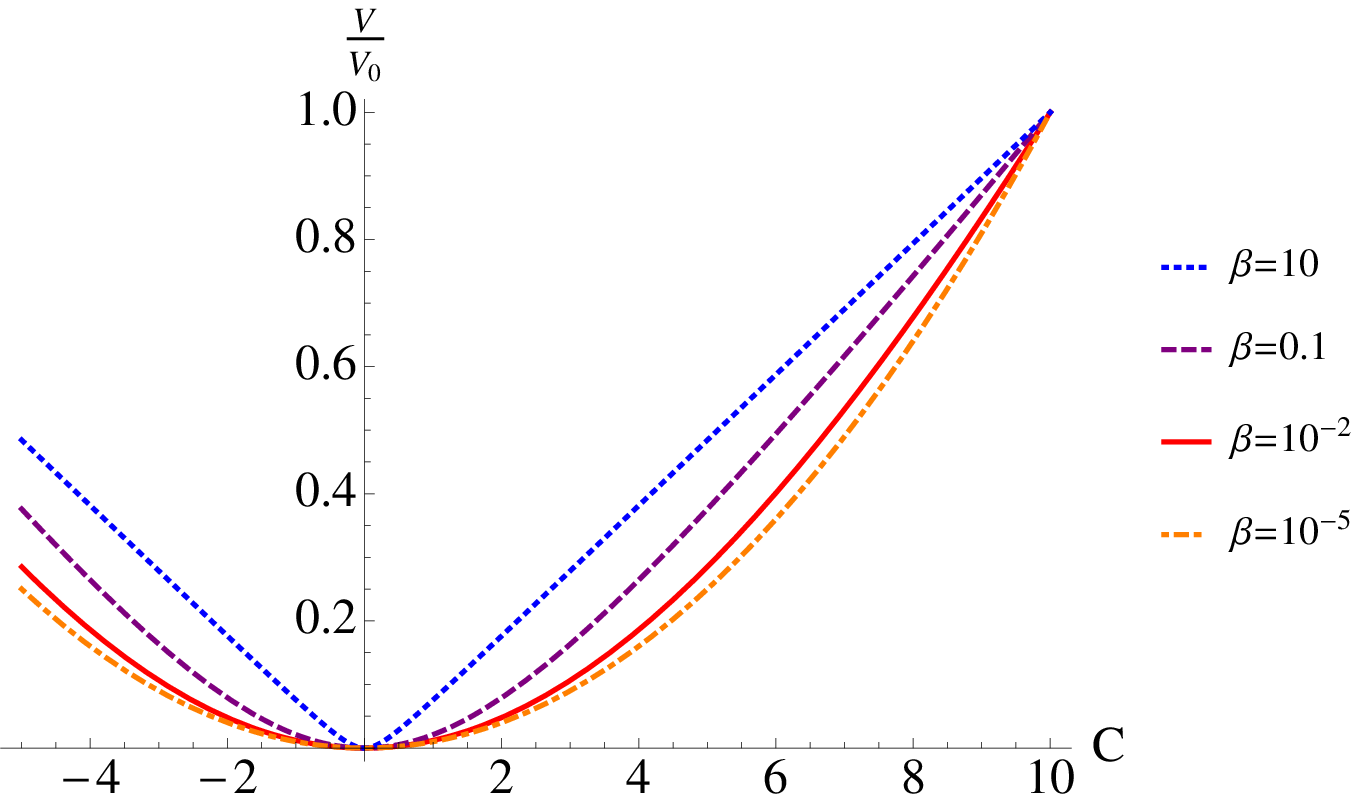}
\caption{The forms of the scalar potential~(\ref{modV3}) with different values of $\beta$ are shown. The potential is normalized at $C=10$ ($V_0\equiv V(C=10)$).}
\label{chaotic.fig}
\end{center}
\end{figurehere}
In Fig.~\ref{chaotic.fig}, the scalar potential asymptotically converges on the linear type potential for larger $\beta$, which is expected from Eq.~(\ref{modV3}). 

Due to the modification of the scalar potential, the predicted values of cosmological parameters are also different from ones of the original quadratic potential if $\beta$ is sufficiently large. We numerically calculated the spectral index $n_s$ and the tensor to scalar ratio $r$ at the horizon exit. We obtained following sets of values for $\beta=10^{-5}$ and $\beta=10$ respectively,
\begin{align}
(n_s,r)=
\begin{cases}
(0.967,0.132)&( \beta=10^{-5})\\
(0.975,0.0666)&( \beta=10),
\end{cases}
\end{align}
where we show the values at $N=60$, and $N$ is the number of e-foldings. In both cases, $g$ should be $\mathcal{O}(10^{-5})$. For $\beta=10^{-5}$, the cut-off scale $M$ is $\mathcal{O}(1)$, and the values of the cosmological parameters are not affected by the DBI corrections, which is compatible with the conclusion of the Ref.~\cite{Ferrara:2013kca}. On the other hand, the predicted parameters are altered by the DBI corrections for $\beta=10$. In this case, $M\sim 10^{-3}$, which is higher than the inflation scale $H\sim 10^{-4}$.

As another example, let us discuss the Starobinsky model, shown in Sec.~\ref{review}. In that case, $J(C)$ is given by $J=-\frac{3}{2}\log( -Ce^C/3)$. Then the Lagrangian of $C$ is given by 
\begin{align}
\mathcal{L}=-\frac{3}{4C^2}\partial_\mu C\partial^\mu C-\frac{g^2}{\beta}\left(\sqrt{1+\frac{9\beta}{4}(1+\frac{1}{C})^2}-1\right).\label{Sc}
\end{align}
We redefine $C$ as $C=-\exp (\sqrt{2/3}\phi)$, and then we can rewrite the action~(\ref{Sc}) as
\begin{align}
\mathcal{L}=-\frac{1}{2}\partial_\mu \phi\partial^\mu \phi-\frac{g^2}{\beta}\left(\sqrt{1+\frac{9\beta}{4}(1- e^{-\sqrt{2/3}\phi})^2}-1\right).\label{Sc2}
\end{align} 
We show the scalar potential in Eq.~(\ref{Sc2}) with various values of $\beta$ in Fig.~\ref{R^2.fig}.
\begin{figurehere}
\begin{center}
\includegraphics[width=85mm]{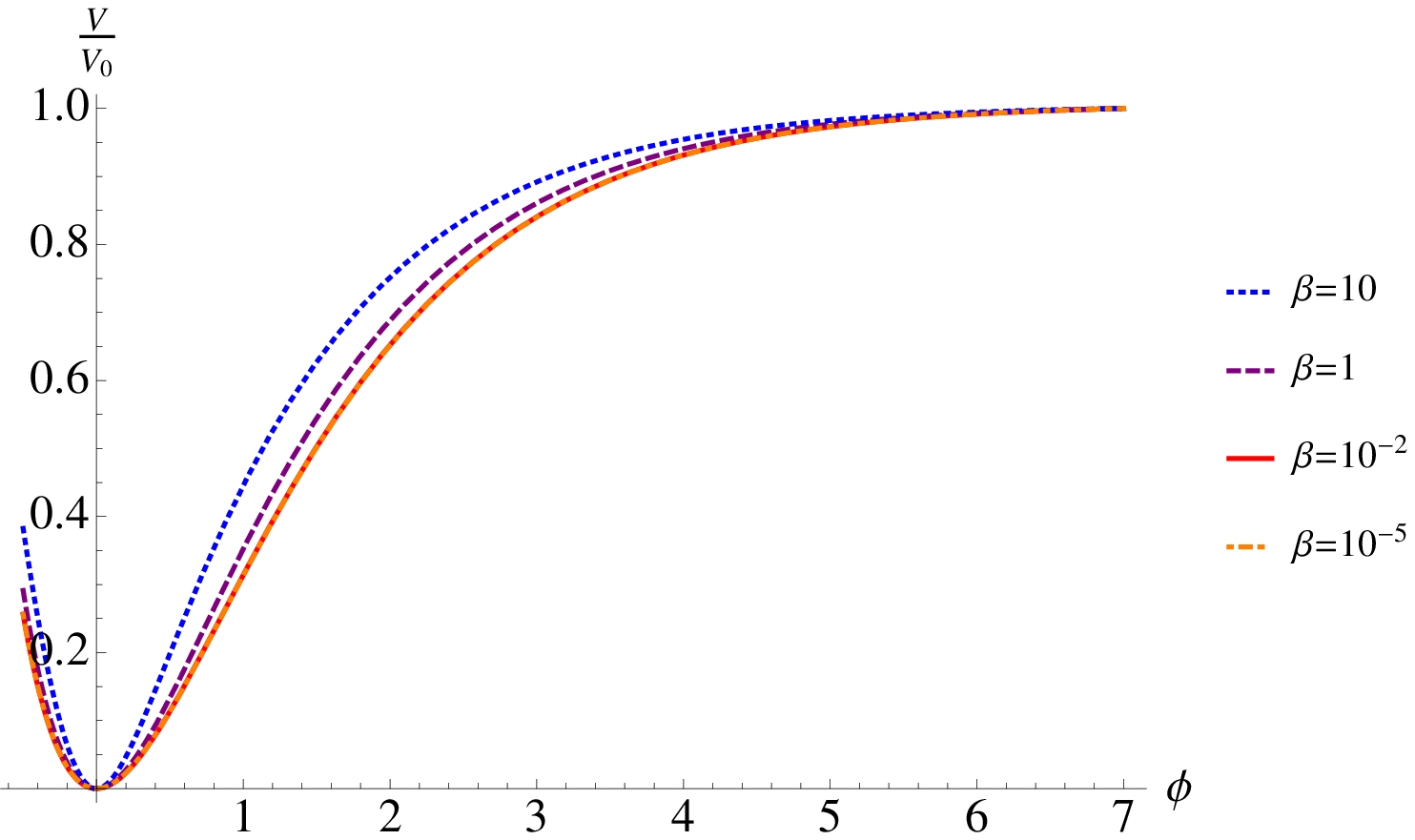}
\caption{The forms of the scalar potential in Eq.~(\ref{Sc2}) with different values of $\beta$ are shown. The potential is normalized at $\phi=7$ ($V_0\equiv V(\phi=7)$).}
\label{R^2.fig}
\end{center}
\end{figurehere}
 As in the previous model, we obtained the following values of $(n_s,r)$ for $\beta=10^{-5}$ and $\beta=10$,
\begin{align}
(n_s,r)=
\begin{cases}
(0.968,0.00296)&( \beta=10^{-5})\\
(0.968,0.00280)&( \beta=10),
\end{cases}\label{R^2nsr}
\end{align}
where we have shown the values at $N=60$. As shown in Fig.~\ref{R^2.fig}, the scalar potential in Eq.~(\ref{Sc2}) is not much affected by the DBI corrections. The reason can be understood as follows. In the region that $\exp(-\sqrt{2/3}\phi) \ll 1$, the original potential in Eq.~(\ref{R^22}) can be approximately written as 
\begin{align}
V=&\frac{9g^2}{8}(1-e^{-\sqrt{\frac{2}{3}}\phi})^2\nonumber\\
\sim& \frac{9g^2}{8}(1-2e^{-\sqrt{\frac{2}{3}}\phi}).
\end{align}
On the other hand, for a sufficiently large $\beta$, the scalar potential in Eq.~(\ref{Sc2}) can be approximated as
\begin{align}
V\sim\frac{g^2}{\beta}\sqrt{\frac{9\beta}{4}(1- e^{-\sqrt{\frac{2}{3}}\phi})^2}=\frac{3g^2}{2\sqrt{\beta}}(1- e^{-\sqrt{\frac{2}{3}}\phi}).
\end{align}
Therefore, the form of the leading order terms is not affected by the DBI corrections, and therefore the predicted cosmological parameters~(\ref{R^2nsr}) are almost unchanged. 
\section{DBI action of $R^2$ model in new minimal SUGRA}\label{DBIR^2}
As we mentioned in Sec.~\ref{intro}, the Starobinsky model in the new minimal SUGRA is dual to the massive vector inflation model with $J=-\frac{3}{2}\log( -Ce^C/3)$ that we treated in Sec.~\ref{review}. We review the Starobinsky model in the new minimal SUGRA in Appendix~\ref{Starobinsky model}.

Since the dual of the new minimal Starobinsky model is described by a massive vector multiplet, it is interesting to construct the DBI-extension of the new minimal Starobinsky model. As we find below, it is dual to the massive DBI action with a special form of $\omega$.

As shown in Eq.~(\ref{ap1}), the Starobinsky model in the new minimal SUGRA consists of a real linear compensator $L_0$ and a real multiplet $V_R=\log (L_0/|{\cal S}|^2)$ where ${\cal S}$ is a chiral multiplet. As is the case with gauge multiplets, we can construct the following DBI type action in the new minimal Starobinsky model,
\begin{align}
S=\left[\frac{3}{2}L_0V_R\right]_D+[-\gamma X(\cW,\bar{\cW},L_0)]_F,\label{sm1}
\end{align}
where $\gamma$ is a real constant, and $X(\cW,\bar{\cW},L_0)$ is a solution of the following equation (see Appendix A in Ref.~\cite{Abe:2015nxa}),
\begin{align}
X=\cW^\alpha(V_R)\cW_\alpha(V_R)-\kappa\Sigma_c(|X|^2L_0^{-2}),\label{sm2}
\end{align}
where $\kappa$ is a positive constant. Since it is difficult to solve Eq.~(\ref{sm2}), we rewrite the action in Eq.~(\ref{sm1}) with a chiral Lagrange multiplier multiplet ${\cal M}$ as
\begin{align}
S=&\left[\frac{3}{2}L_0V_R\right]_D+[-\gamma X]_F\nonumber\\
&+[2{\cal M}(\cW^2(V_R)-\kappa\Sigma_c(|X|^2L_0^{-2})-X)]_F.\label{sm3}
\end{align}
After the gauge fixings of $V_R$ and the superconformal symmetry, the bosonic Lagrangian of the action~(\ref{sm3}) becomes
\begin{align}
\mathcal{L}|_B=&\frac{1}{2}R+\frac{3}{4}B_\mu B^\mu-\frac{3}{2}A_\mu^{(R)}B^\mu+4\kappa\lambda|F^X|^2-4\lambda D_{(R)}^2\nonumber\\
&+2\lambda \cF_{\mu\nu}^{(R)}\cF^{(R)\mu\nu}-2i\chi \cF_{\mu\nu}^{(R)}\tilde{\cF}^{(R)\mu\nu}\nonumber\\
&-(\gamma+2{\cal M})F^X-(\gamma+2\bar{\cal M})\bar{F}^{\bar{X}},\label{sm4}
\end{align}
where $F^X$ is the F-term of $X$, $\lambda\equiv {\rm Re}{\cal M}$, $\chi\equiv{\rm Im}{\cal M}$, $D_{(R)} \equiv (R+3B_\mu B^\mu / 2)/3$, $A_\mu^{(R)}$ is the vector component of $V_R$, $B_\mu$ is an auxiliary vector component of $L_0$, and $\cF_{\mu\nu}^{(R)}=\partial_\mu A^{(R)}_\nu-\partial_\nu A^{(R)}_\mu$. Integrating out $F^X$, $\lambda$, $\chi$, we obtain the following action.
\begin{align}
\mathcal{L}|_B=&\frac{1}{2}R+\frac{3}{4}B_\mu B^\mu-\frac{3}{2}A_\mu^{(R)}B^\mu-\frac{\gamma}{\kappa}\nonumber\\
+&\frac{\gamma}{\kappa}\sqrt{1+4\kappa D_{(R)}^2-2\kappa \cF_{\mu\nu}^{(R)}\cF^{(R)\mu\nu}+\kappa^2(\cF_{\mu\nu}^{(R)}\tilde{\cF}^{(R)\mu\nu})^2}\nonumber\\
=&\frac{1}{2}R+\frac{3}{4}B_\mu B^\mu-\frac{3}{2}A_\mu^{(R)}B^\mu-\frac{\gamma}{\kappa}\nonumber\\
+&\frac{\gamma}{\kappa}\sqrt{4\kappa D_{(R)}^2-{\rm det}(\eta_{ab}+\sqrt{\kappa}\cF_{ab})}.\label{sm5}
\end{align}
The action~(\ref{sm5}) has the DBI type structure with respect to the vector $A_\mu^{(R)}$, but it also includes $D_{(R)}^2=(R+3B_\mu B^\mu/2)^2$ in the square root, which yields the higher curvature correction~\cite{Deser:1998rj,Gates:2001ff}.

Next, let us discuss the dual action of~(\ref{sm5}). As performed in Appendix~\ref{Starobinsky model}, we can derive the dual action by using a real linear multiplier multiplet $U$ and a gauge multiplet $V$.
Let us consider the following action.
\begin{align}
S=&\left[\frac{3}{2}L_0V_R\right]_D+[-\gamma X]_F\nonumber\\
&+[2{\cal M}(\cW^2(V)-\kappa\Sigma_c(|X|^2L_0^{-2})-X)]_F\nonumber\\
&+[U(V-V_R)]_D.\label{sm6}
\end{align}
The equation of motion of $U$ gives a solution with respect to $V$ shown in Eq.~(\ref{ap5}), which reproduces the action~(\ref{sm3}). On the other hand, if we solve it with respect to $L_0$, which is shown in Eq.~(\ref{ap6}), the action~(\ref{sm6}) becomes
\begin{align}
S=&\left[\frac{3}{2}|{\cal S}|^2(\Lambda+\bar{\Lambda}+V)\exp (\Lambda+\bar{\Lambda}+V)\right]_D+[-\gamma X]_F\nonumber\\
&+[2{\cal M}(\cW^2(V)-\kappa\Sigma_c(|X|^2|{\cal S}|^{-2}e^{-2(\Lambda+\bar{\Lambda}+V)})-X)]_F.
\end{align}
By the field redefinitions~$\Lambda\to\Lambda/2,\ V\to gV/2,\ {\cal S}\to S_0/\sqrt{3},\ {\cal M}\to 4{\cal M}/g^2, X\to g^2X/4$, we obtain
\begin{align}
S=&\left[\frac{a}{2}|S_0|^2Ce^C\right]_D+\left[-\frac{g^2\gamma}{4} X\right]_F\nonumber\\
&+[2{\cal M}(\cW^2(V)-\frac{3\kappa g^2}{4}\Sigma_c(|X|^2|S_0|^{-2}e^{-2C})-X)]_F,\label{sm7}
\end{align}
where $C=(\Lambda+\bar{\Lambda}+gV)/2$.  
We choose parameters $\gamma$ and $\kappa$ as $\gamma=g^{-2}$, $\kappa=3g^{-2}M^{-4}$, and then the action~(\ref{sm7}) becomes
\begin{align}
S=&\int d^4x\sqrt{-g}\Biggl[\frac{1}{2}R-\frac{1}{2}J''(C)\partial_\mu C\partial^\mu C-\frac{g^2}{2}J''A_\mu A^\mu\nonumber\\
+&\int d^4x \frac{M^4}{C^2}\left(1-\sqrt{P}\sqrt{-{\rm det}\left(g_{\mu\nu}+\frac{C}{M^2}\cF_{\mu\nu}\right)}\right),\label{sm8}
\end{align}
where
\begin{align}
P=1+\frac{g^2C^2(J'(C))^2}{M^4}.
\end{align}
The action~(\ref{sm8}) is a special case of the DBI-extended Starobinsky model, i.e.,
Eq.~(\ref{DBI3}) with $J=-\frac{3}{2}\log(-C e^C/3)$.
In fact, we can reproduce the action~(\ref{sm8}) from the latter by choosing $\omega$ as
\begin{align}
 \omega = \frac{9e^{-2C}}{4M^4}.
 \end{align}
Note that this choice is different from the one in the previous sections:
\begin{align}
\omega= \frac{e^{\frac{4}{3}J}}{4M^4} = \frac{9e^{-2C}}{4M^4C^2},\label{Romega}
\end{align}
which is chosen so that the coefficient of ${\cal F}_{\mu\nu}$ in the square root in (\ref{DBI3}) becomes a constant,
just like the ordinary DBI action.
Namely, although Eq.(\ref{DBI3}) with $J=-\frac{3}{2}\log(-Ce^C/3)$ and with (\ref{Romega}) is the DBI-extension of
the dual action to the Starobinsky model, it is not the dual action to the DBI-extension of
the (new minimal) Starobinsky model, which is given by (\ref{sm8}).

In terms of the canonical scalar field $\phi=\sqrt{3/2} \log(-C)$, we can obtain the following scalar potential.
\begin{align}
V=\frac{g^2}{\beta}e^{-2\sqrt{\frac{2}{3}}\phi}\left(\sqrt{1+\frac{9\beta e^{2\sqrt{\frac{2}{3}}\phi}}{4}(1-e^{-\sqrt{\frac{2}{3}\phi}})^2}-1\right),\label{sm9}
\end{align} 
where $\beta=g^2/M^4$. For comparison, we show the scalar potential~(\ref{sm9}) with different values of $\beta$ and the original one in the Starobinsky model~(\ref{R^22}) in Fig.~\ref{SDBI}.
\begin{figurehere}
\begin{center}
\includegraphics[width=85mm]{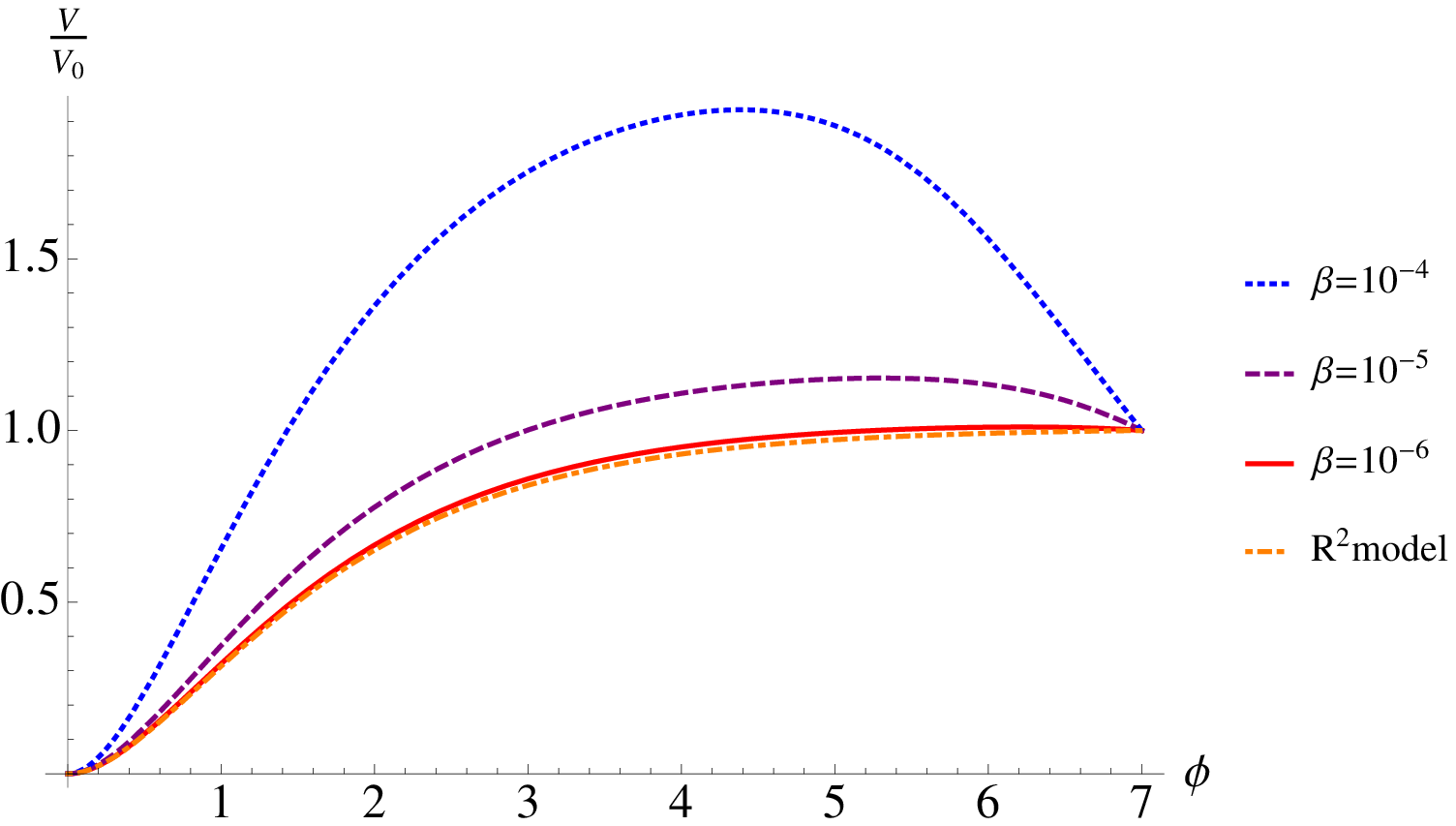}
\caption{The forms of the scalar potential~(\ref{sm9}) with different values of $\beta$ and the original one in Eq.~(\ref{R^22}) are shown. The potential is normalized at $\phi=7$ ($V_0\equiv V(\phi=7)$).}
\label{SDBI}
\end{center}
\end{figurehere}
From Fig.~\ref{SDBI}, we find that the scalar potential in Eq.~(\ref{sm9}) changes drastically even with a small $\beta$, unlikely to the one in Eq.~(\ref{Sc2}) discussed in Sec.\ref{inflation}.
We can understand this property as follows. For $A_\mu^{(R)}=0$, the auxiliary field $B_\mu$ is integrated out and we obtain $B_\mu=0$, then the Lagrangian~(\ref{sm5}) becomes 
\begin{align}
\mathcal{L}=\frac{1}{2}R+\frac{g^2}{3\beta}(\sqrt{1+12\beta R^2}-1).
\end{align}
If $\beta R^2\gg 1$, the structure $\sim R+R^2$ does not appear, that means the Starobinsky model appears as the low curvature (compared with  $1/\sqrt{\beta}$) approximation in the DBI-extension of a massive vector action dual to the one in new minimal SUGRA. In this case, the conformal symmetry, which the term $R^2$ has, is severely broken. Thus, the inflaton potential loses its flat profile. We remark that, from the viewpoint of superconformal formulation, the essential difference between the action (\ref{DBI2}) in section~\ref{massiveDBI} and the one (\ref{sm8}) in this section is the choice of the function $\omega$ in Eq.~(\ref{cond}). 
\section{Conclusion}\label{conclusion}
We have constructed the DBI action of a massive vector multiplet in 4D ${\cal N}=1$ SUGRA, and also shown how the DBI corrections affect the inflationary potential in the massive vector inflation model. Such an extension may be required to realize the model in superstring theory if the action is an effective action of D-brane. As shown in Sec.~\ref{massiveDBI}, the DBI action of a massive vector multiplet is obtained from the massless one coupled to the Stueckelberg chiral multiplet, which is  based on the result in our previous work~\cite{Abe:2015nxa}. As a nontrivial consequence of the DBI-extension, the D-term potential is modified. 

It is a common expectation that the higher order corrections ruin the flatness of the scalar potential. However, we have found that the modified scalar potential is flatter than the one before the DBI-extension, as shown in Sec.~\ref{inflation}. We have also shown the concrete examples, the quadratic chaotic inflation and the Starobinky inflation. Especially for the former case, the scalar potential is drastically modified into the linear one, even with the cut-off scale $M$ larger than the Hubble scale during inflation. Such an interesting feature of the DBI action may be important for inflation models in SUGRA. The feature is also favored by the latest results from the CMB observation by the Planck satellite~\cite{Ade:2015lrj}, because the DBI correction reduces the value of $r$ (see examples in Sec.\ref{inflation}). It is also worth to note that the functions $J(C)$ and $\omega(\Lambda,\bar{\Lambda})$ are not so arbitrary, restricted by geometries and symmetries of the background, if it is realized in superstring theory. If the simplest case given in Eq.~(\ref{DBI2}) is realized, the DBI correction discussed in this work always flatten the potential as shown below Eq.~(\ref{modV2}).

We have also discussed the DBI-extension of the Starobinsky model in the new minimal SUGRA. The action~(\ref{sm5}) is a possible extension of the new minimal SUGRA, which has higher order terms of $R$ without ghosts. As shown in Sec.~\ref{DBIR^2}, the action is dual to a special class of the massive DBI action, which is different from the simple massive DBI action~(\ref{DBI2}). In that case, the scalar potential is highly affected by the DBI corrections and it loses the flat profile because of the breaking of the conformal symmetry, which the original Strarobinsky model has during inflation. 
\section*{Acknowledgments}
H.A., Y.S. and Y.Y. are supported in part by Grant-in-Aid for Young Scientists (B) 
(No. 25800158),
Grant-in-Aid for Scientific Research (C) (No.25400283),    
and Research Fellowships for Young Scientists (No.26-4236), 
which are from Japan Society for the Promotion of Science, respectively. 
\begin{appendix}
\section{The Starobinsky model in new minimal SUGRA}\label{Starobinsky model}
We briefly review the Starobinsky model in the new minimal SUGRA~\cite{Farakos:2013cqa}. From the conformal SUGRA viewpoint, the new minimal SUGRA can be regarded as conformal SUGRA with a real linear compensator multiplet denoted by $L_0$.

Supermultiplets in conformal SUGRA are characterized by the Weyl and the chiral weights denoted by $w$ and $n$ respectively, which are the charges for the dilatation and U$(1)_A$.\footnote{The dilatation and U$(1)_A$ are parts of the 4 dimensional ${\cal N}=1$ superconformal algebra.} $L_0$ has the weights $(w,n)=(2,0)$.
 
The Starobinsky action in the new minimal SUGRA is given by
\begin{align}
S=\left[\frac{3}{2}L_0V_R\right]_D+[-\gamma\cW^\alpha(V_R)\cW_{\alpha}(V_R)]_F,\label{ap1}
\end{align}
where $V_R=\log (L_0/|{\cal S}|^2)$, and ${\cal S}$ is a chiral multiplet with $(w,n)=(1,1)$, and $\gamma$ is a positive constant.  It is worth noting that a linear multiplet has a property~$[L(\Lambda+\bar{\Lambda})]_D=0$, where $\Lambda$ is a chiral multiplet with $(w,n)=(0,0)$. Therefore, the action~(\ref{ap1}) is invariant under the transformation~
${\cal S}\to {\cal S} e^{\Sigma}$.
Under this transformation, $V_R$ transforms as $V_R\to V_R-\Sigma-\bar{\Sigma}$ like a gauge multiplet shown in Eq.~(\ref{trans}). Therefore, we can perform the ``gauge" transformation to remove ${\cal S}$ from the physical action, and we choose the gauge in which ${\cal S}=1$. After the superconformal gauge fixings, the bosonic part of the Lagrangian $\mathcal{L}$ in Eq.~(\ref{ap1}) can be calculated as
\begin{align} 
\mathcal{L}|_B=&\frac{1}{2}R+\frac{2\gamma}{9}R^2-\gamma \cF^{(R)}_{\mu\nu}\cF^{(R)\mu\nu}-\frac{3}{2}A_\mu^{(R)}B^{\mu}\nonumber\\
&+\left(\frac{3}{4}+\frac{2\gamma}{3}R\right)B_\mu B^{\mu}+\gamma (B_\mu B^\mu)^2,\label{ap4}
\end{align}
where $A_\mu^{(R)}$ is the vector component of $V_R$, $\cF^{(R)}_{\mu\nu}=\partial_\mu A^{(R)}_{\nu}-\partial_\nu A^{(R)}_{\mu}$, and $B_\mu$ is an auxiliary vector component of $L_0$. On the trajectory~$A_\mu^{(R)}=0$, we can integrate $B_\mu$ out, and obtain the usual Starobinsky Lagrangian~$\mathcal{L}=R/2+2\gamma R^2/9$.

Since $V_R$ transforms like a U(1) gauge multiplet, it is natural to think that the action~(\ref{ap1}) has a dual form with a gauge multiplet. Indeed, the action~(\ref{ap1}) can be rewritten as
\begin{align}
S=&\left[\frac{3}{2}L_0V_R\right]_D+[-\gamma\cW^\alpha(V)\cW_{\alpha}(V)]_F\nonumber\\
&+[U(V-V_R)]_D\label{ap2}
\end{align}
where $U$ is a real linear multiplet, and $V$ is a gauge multiplet. By varying $U$, we obtain $V=V_R-\Lambda -\bar{\Lambda}$ where $\Lambda$ is a chiral multiplet. Substituting this into Eq.~(\ref{ap2}), we can reproduce the action~(\ref{ap1}). On the other hand, we can also solve the equation of motion of $U$ with respect to $L_0$ as follows.
\begin{align}
V_R=\log\left(\frac{L_0}{|{\cal S}|^2}\right)=(\Lambda+\bar{\Lambda}+V),\label{ap5}
\end{align} 
equivalently,
\begin{align}
L_0=|{\cal S}|^2\exp \left(\Lambda+\bar{\Lambda}+V\right).\label{ap6}
\end{align}
Thus, we obtain the following dual action,
\begin{align}
S_{\rm dual}=&\left[\frac{3}{2}|{\cal S}|^2(\Lambda+\bar{\Lambda}+V)\exp (\Lambda+\bar{\Lambda}+V)\right]_D\nonumber\\
&+[-\gamma\cW^\alpha(V)\cW_{\alpha}(V)]_F.
\end{align}
After field redefinitions~$\Lambda\to \Lambda/2,\ V\to gV/2, \ {\cal S}\to S_0/\sqrt{3}$, we obtain
\begin{align}
S_{\rm dual}=&\left[\frac{1}{2}|S_0|^2\left(\frac{1}{2}(\Lambda+\bar{\Lambda}+gV)\right)\exp \left(\frac{1}{2}(\Lambda+\bar{\Lambda}+gV)\right)\right]_D\nonumber\\
&-\frac{g^2\gamma}{4}[\cW^\alpha(V)\cW_{\alpha}(V)]_F.\label{ap3}
\end{align}
For $\gamma=g^{-2}$, the action~(\ref{ap3}) reproduces one in Eq.~(\ref{S1}) with $\Phi(C)=Ce^C$.
\end{appendix}

\end{multicols}

\begin{thebibliography}{99}
\bibitem{Guth:1980zm} 
  A.~H.~Guth,
  ``The Inflationary Universe: A Possible Solution to the Horizon and Flatness Problems,''
  Phys.\ Rev.\ D {\bf 23}, 347 (1981),
  
\bibitem{Starobinsky:1980te} 
  A.~A.~Starobinsky,
  ``A New Type of Isotropic Cosmological Models Without Singularity,''
  Phys.\ Lett.\ B {\bf 91}, 99 (1980).
\bibitem{Sato:1980yn} 
  K.~Sato,
  ``First Order Phase Transition of a Vacuum and Expansion of the Universe,''
  Mon.\ Not.\ Roy.\ Astron.\ Soc.\  {\bf 195}, 467 (1981).
\bibitem{Linde:1983gd} 
  A.~D.~Linde,
  ``Chaotic Inflation,''
  Phys.\ Lett.\ B {\bf 129}, 177 (1983).
\bibitem{Kawasaki:2000yn} 
  M.~Kawasaki, M.~Yamaguchi and T.~Yanagida,
  ``Natural chaotic inflation in supergravity,''
  Phys.\ Rev.\ Lett.\  {\bf 85}, 3572 (2000)
  [hep-ph/0004243].
\bibitem{Kallosh:2010xz} 
  R.~Kallosh, A.~Linde and T.~Rube,
  ``General inflaton potentials in supergravity,''
  Phys.\ Rev.\ D {\bf 83}, 043507 (2011)
  [arXiv:1011.5945 [hep-th]].
\bibitem{Demozzi:2010aj} 
  V.~Demozzi, A.~Linde and V.~Mukhanov,
  ``Supercurvaton,''
  JCAP {\bf 1104}, 013 (2011)
  [arXiv:1012.0549 [hep-th]].
\bibitem{Goncharov:1983mw} 
  A.~B.~Goncharov and A.~D.~Linde,
  ``Chaotic Inflation in Supergravity,''
  Phys.\ Lett.\ B {\bf 139}, 27 (1984),

  A.~S.~Goncharov and A.~D.~Linde,
  ``Chaotic Inflation Of The Universe In Supergravity,''
  Sov.\ Phys.\ JETP {\bf 59}, 930 (1984)
  [Zh.\ Eksp.\ Teor.\ Fiz.\  {\bf 86}, 1594 (1984)].
  
\bibitem{Ketov:2014qha} 
  S.~V.~Ketov and T.~Terada,
  ``Inflation in Supergravity with a Single Chiral Superfield,''
  Phys.\ Lett.\ B {\bf 736}, 272 (2014)
  [arXiv:1406.0252 [hep-th]],
  
  S.~V.~Ketov and T.~Terada,
  ``Generic Scalar Potentials for Inflation in Supergravity with a Single Chiral Superfield,''
  JHEP {\bf 1412}, 062 (2014)
  [arXiv:1408.6524 [hep-th]],
  
  D.~Roest and M.~Scalisi,
  ``Cosmological Attractors from $\alpha$-Scale Supergravity,''
  arXiv:1503.07909 [hep-th],
  
  A.~Linde,
  ``Single-field $\alpha$-attractors,'' arXiv:1504.00663 [hep-th].
\bibitem{Kawano:2007gg} 
  T.~Kawano,
  ``Chaotic D-term inflation,''
  Prog.\ Theor.\ Phys.\  {\bf 120}, 793 (2008)
  [arXiv:0712.2351 [hep-th]].
\bibitem{Ferrara:2013rsa} 
  S.~Ferrara, R.~Kallosh, A.~Linde and M.~Porrati,
  ``Minimal Supergravity Models of Inflation,''
  Phys.\ Rev.\ D {\bf 88}, no. 8, 085038 (2013)
  [arXiv:1307.7696 [hep-th]].
\bibitem{Farakos:2014gba} 
  F.~Farakos and R.~von Unge,
  ``Naturalness and Chaotic Inflation in Supergravity from Massive Vector Multiplets,''
  JHEP {\bf 1408}, 168 (2014)
  [arXiv:1404.3739 [hep-th]].
  \bibitem{VanProeyen:1979ks} 
  A.~Van Proeyen,
  ``Massive Vector Multiplets in Supergravity,''
  Nucl.\ Phys.\ B {\bf 162}, 376 (1980).
\bibitem{Mukhi:1979wc} 
  S.~Mukhi,
  ``Massive Vector Multiplet Coupled To Supergravity,''
  Phys.\ Rev.\ D {\bf 20}, 1839 (1979).
\bibitem{Born:1934gh} 
  M.~Born and L.~Infeld,
  ``Foundations of the new field theory,''
  Proc.\ Roy.\ Soc.\ Lond.\ A {\bf 144}, 425 (1934).
\bibitem{Dirac:1962iy} 
  P.~A.~M.~Dirac,
  ``An Extensible model of the electron,''
  Proc.\ Roy.\ Soc.\ Lond.\ A {\bf 268}, 57 (1962).
\bibitem{Tseytlin:1999dj} 
  A.~A.~Tseytlin,
  In *Shifman, M.A. (ed.): The many faces of the superworld* 417-452
  [hep-th/9908105].
  \bibitem{Cecotti:1986gb} 
  S.~Cecotti and S.~Ferrara,
  Phys.\ Lett.\ B {\bf 187}, 335 (1987).
\bibitem{Bagger:1996wp} 
  J.~Bagger and A.~Galperin,
  Phys.\ Rev.\ D {\bf 55}, 1091 (1997)
  [hep-th/9608177].
\bibitem{Rocek:1997hi} 
  M.~Rocek and A.~A.~Tseytlin,
  Phys.\ Rev.\ D {\bf 59}, 106001 (1999)
  [hep-th/9811232].
\bibitem{Ketov:1998ku} 
  S.~V.~Ketov,
  Mod.\ Phys.\ Lett.\ A {\bf 14}, 501 (1999)
  [hep-th/9809121].
\bibitem{Ketov:2001dq} 
  S.~V.~Ketov,
  hep-th/0108189.
\bibitem{Kuzenko:2002vk} 
  S.~M.~Kuzenko and S.~A.~McCarthy,
  ``Nonlinear selfduality and supergravity,''
  JHEP {\bf 0302}, 038 (2003)
  [hep-th/0212039].
\bibitem{Kuzenko:2005wh} 
  S.~M.~Kuzenko and S.~A.~McCarthy,
  ``On the component structure of ${\cal N}=1$  supersymmetric nonlinear electrodynamics,''
  JHEP {\bf 0505}, 012 (2005)
  [hep-th/0501172].
\cite{Ketov:2003gr}
\bibitem{Ketov:2003gr} 
  S.~V.~Ketov,
  Mod.\ Phys.\ Lett.\ A {\bf 18}, 1887 (2003)
  [hep-th/0304002].
\bibitem{Abe:2015nxa} 
  H.~Abe, Y.~Sakamura and Y.~Yamada,
  ``Matter coupled Dirac-Born-Infeld action in 4-dimensional N=1 conformal supergravity,''
  arXiv:1504.01221 [hep-th].
\bibitem{Cecotti:1987qe} 
  S.~Cecotti, S.~Ferrara, M.~Porrati and S.~Sabharwal,
  ``New Minimal Higher Derivative Supergravity Coupled To Matter,''
  Nucl.\ Phys.\ B {\bf 306}, 160 (1988).
 
  \bibitem{Kaku:1978nz} 
  M.~Kaku, P.~K.~Townsend and P.~van Nieuwenhuizen,
  ``Properties of Conformal Supergravity,''
  Phys.\ Rev.\ D {\bf 17}, 3179 (1978),
  
  M.~Kaku and P.~K.~Townsend,
  ``Poincare Supergravity As Broken Superconformal Gravity,''
  Phys.\ Lett.\ B {\bf 76}, 54 (1978),
  
  P.~K.~Townsend and P.~van Nieuwenhuizen,
  ``Simplifications of Conformal Supergravity,''
  Phys.\ Rev.\ D {\bf 19}, 3166 (1979).
\bibitem{Kugo:1982cu} 
  T.~Kugo and S.~Uehara,
  ``Conformal and Poincare Tensor Calculi in $N=1$ Supergravity,''
  Nucl.\ Phys.\ B {\bf 226}, 49 (1983).
\bibitem{Cecotti:1987qr} 
  S.~Cecotti, S.~Ferrara and L.~Girardello,
  ``Massive Vector Multiplets From Superstrings,''
  Nucl.\ Phys.\ B {\bf 294}, 537 (1987).
    \bibitem{Kugo:1982mr} 
  T.~Kugo and S.~Uehara,
  ``Improved Superconformal Gauge Conditions in the $N=1$ Supergravity {Yang-Mills} Matter System,''
  Nucl.\ Phys.\ B {\bf 222}, 125 (1983).
\bibitem{Farakos:2013cqa} 
  F.~Farakos, A.~Kehagias and A.~Riotto,
  ``On the Starobinsky Model of Inflation from Supergravity,''
  Nucl.\ Phys.\ B {\bf 876}, 187 (2013)
  [arXiv:1307.1137].
\bibitem{Green:1984sg} 
  M.~B.~Green and J.~H.~Schwarz,
  ``Anomaly Cancellation in Supersymmetric D=10 Gauge Theory and Superstring Theory,''
  Phys.\ Lett.\ B {\bf 149}, 117 (1984).
\bibitem{Lopes Cardoso:1991zt} 
  G.~Lopes Cardoso and B.~A.~Ovrut,
  ``A Green-Schwarz mechanism for D = 4, $N=1$  supergravity anomalies,''
  Nucl.\ Phys.\ B {\bf 369}, 351 (1992).

\bibitem{Ferrara:2013kca} 
  S.~Ferrara, R.~Kallosh, A.~Linde and M.~Porrati,
  ``Higher Order Corrections in Minimal Supergravity Models of Inflation,''
  JCAP {\bf 1311}, 046 (2013)
  [arXiv:1309.1085].
\bibitem{Deser:1998rj} 
  S.~Deser and G.~W.~Gibbons,
  Class.\ Quant.\ Grav.\  {\bf 15}, L35 (1998)
  [hep-th/9803049].
\bibitem{Gates:2001ff} 
  S.~J.~Gates, Jr. and S.~V.~Ketov,
  Class.\ Quant.\ Grav.\  {\bf 18}, 3561 (2001)
  [hep-th/0104223].
\bibitem{Ade:2015lrj} 
  P.~A.~R.~Ade {\it et al.}  [Planck Collaboration],
  arXiv:1502.02114 [astro-ph.CO].
\end{thebibliography}
\end{document}